\documentstyle[12pt]{article}

\newenvironment{figcap}{
     \newpage\pagestyle{empty}\thispagestyle{empty}
     \section{\ }\medskip
    \begin{list}{{\large \bf Fig.~\arabic{myeqsno}}}{
      \usecounter{myeqsno}\setlength{\rightmargin}{\leftmargin}}}{
    \end{list}}
\newcounter{myeqsno}
\def\av#1{\langle #1 \rangle}

\def\ta{t_{\rm age}}
\def\mtrm{M_{\rm TRM}}

\begin{document}
\begin{titlepage}
\title{Thermocycling experiments with the three-dimensional Ising spin
glass model}
\author{Heiko Rieger\thanks{e-mail: rieger@thp.uni-koeln.de}\\
\mbox{$\quad$}\\
Institut f\"ur Theoretische Physik\\
Universit\"at zu K\"oln\\
50937 K\"oln,Germany}
\date{}
\maketitle
\thispagestyle{empty}
\begin{abstract}
\normalsize
\noindent
A characteristic feature of the non--equilibrium dynamics of real
spin glasses at low temperatures are strong aging effects. These
phenomena can be manipulated by changing the external parameters in
various ways: a thermo-cycling experiment consists for instance of a
short heat pulse during the waiting time, by which the relaxation might
be strongly affected. Results of numerical experiments of
this kind, performed via Monte Carlo simulations of the three dimensional
Ising spin glass model, are presented. The theoretical implications
are discussed and the scenario found is compared with the experimental
situation.
\end{abstract}
\vspace*{1cm}
\end{titlepage}

\section{Introduction}
The non-equilibrium dynamics of spin glasses has been a major focus of
research activity since the seminal work of Lundgren et al.\
\cite{EXP1} of 1983. Two years later the so called aging phenomena
of spin glasses became manifest, when experimentalists realized
\cite{EXP2} that magnetic properties of spin glasses depend strongly
on the time they spent below the freezing temperature $T_g$.
These effects are a consequence of the extremely slow dynamics of
spin glasses at low temperatures, regardless of the existence of
an equilibrium phase transition (for an overview of  their equilibrium
properties see \cite {BY}).

Following the arguments of the droplet theory
\cite{FiHu} one imagines the dynamical process at low temperatures to
be governed by a growth of domains, which will reach some typical size after
a certain waiting time. This length/time scale becomes manifest in a
crossover observable e.g.\ in the thermo-remanent magnetization decay
(see also \cite{KoHi}). On the other hand, inspired by Parisi's solution
of the mean--field model of spin glasses \cite{Pa}, it has been proposed
that the many metastable states existing in the rough free energy landscape
of spin glasses might be organized in a hierarchical way\cite{Leder}.
Dynamics then explores states with decreasing free energy, and the escape from
these valleys becomes harder, which means that it takes longer time. In
this way the depth of the valley reached during the waiting time
becomes manifest again as a crossover in observable quantities (see
also \cite{Bouch}).

Thermo-cycling experiments consist of two temperature changes within the
spin glass phase \cite{Refregier}: either a short heat pulse is applied to
the spin glass during the waiting time after which the relaxation
of e.g.\ the thermo-remanent magnetization is measured, or a short
negative temperature cycle is performed. It has been pointed out
\cite{Lefloch} that this kind of experiments can discriminate between
the droplet picture and the hierarchical picture.

Within the hierarchical picture a
heat pulse experiment and a negative temperature
cycling experiment should have an asymmetric outcome: The hierarchical
organization of the metastable states depends on temperature in such a way
that the free--energy--valleys split into several, steeper valleys with
decreasing temperature. Here an escape becomes even more difficult,
resulting in a freezing of the aging process during a negative temperature
cycle. Upon heating several valleys melt together, thus facilitating
the escape. After completion of the cycle the system is again in a
narrow valley with a higher free energy, similar to the beginning of
the experiment and thus restarting the aging process.
Experiments using the insulating spin glass
CdCr$_{1.7}$In$_{0.3}$S$_4$ \cite{Refregier,Lefloch}
can be nicely interpreted within this picture.

On the other hand within the
droplet theory a heat pulse experiment and a negative temperature
cycling experiment should have qualitatively a symmetric outcome:
in both cases the domains that have grown so far should be reduced in
size (or even destroyed) by the temperature cycle, and thus reinitialize
the aging process. This is a consequence of the fact that spatial
correlations among equilibrium spin glass states at different temperatures
below $T_g$ are short ranged with a typical length scale decresing
rapidly with increasing temperature difference \cite{FiHu,KoHi,BM}.
Experiments in Cu(Mn) spin glasses \cite{Granberg} and Cu(Mn) spin glass
films \cite{Mattson} have reported a scenario that speak in favor of this
interpretation.

Besides these two phenomenological approaches cited above some success
has been made very recently in the analytical and numerical investigation of
microscopic mean-field models of aging in spin glasses
\cite{Crisanti,CuKu,Ritort} or
related, simplified models \cite{PaMa,MeFr}. However, up to now the study of
microscopic models of aging phenomena in two-- or three--dimensional model
of spin glasses had to be performed exclusively via numerical simulations
\cite{JOA,Rieger}. Here not only a qualitative agreement among simulation
results and simple aging--experiment (without temperature cycle) could be
established, but also numerical values for various exponents determining
the functional form of the decay of
the thermo-remanent magnetization have been shown in \cite{Rieger} to be
the same in certain spin glasses
(Fe$_x$Ni$_{(1-x)}$)$_{75}$P$_{16}$B$_6$Al$_3$
and Fe$_{0.5}$Mn$_{0.5}$TiO$_3$) and the three--dimensional
Edwards--Anderson model with a heat--bath dynamics.

One might ask, whether temperature cycling in Monte--Carlo simulations
will also show such a remarkable concurrence with the experimental scenario.
In this paper we present the results of such simulations. In the next
section we introduce the model, give some numerical details and define
the quantities that have been calculated. In section 3 we present the
results, which are discussed in section 4.

\section{Numerical Procedure}

The system under consideration is the same as in \cite{Rieger}, the
three--dimensional Ising spin glass model with nearest neighbor interactions
and a discrete ($\pm J$) bond distribution. Its Hamiltonian is
\begin{equation}
{\cal H} = -\sum_{\langle ij\rangle} J_{ij}\sigma_i\sigma_j
-h\sum_i\sigma_i\;,\label{hamilt}
\end{equation}
where the spins $\sigma_i=\pm1$ occupy the sites of a $L\times L\times L$
simple cubic lattice with periodic boundary conditions and the random
nearest neighbor interactions $J_{ij}$ take on the values $+1$ or $-1$
with probability $1/2$. The quantity $h$ is an external magnetic field.
The dynamics is the so called Metropolis algorithm, defined by the
probability for single spin flips
\begin{equation}
w(\sigma_i\rightarrow-\sigma_i) = {\rm min}\{1,\,\exp(-\Delta E/T)\}\;,
\label{dyn}
\end{equation}
where $\Delta E$ is the energy difference between the old state and the
new state in which the spin at site $i$ is flipped. The most efficient
implementation of this algorithm on a Cray-YMP has been used (see
\cite{mspin} for details).

To mimic the temperature cycling experiments
\cite{Refregier,Lefloch,Granberg,Mattson} the following procedure
has been applied: First the system is initialized in a random state
at a temperature $T<T_g\approx1.2$ ($T_g$ is the freezing or spin glass
transition temperature of the model (\ref{hamilt}), see \cite{eqlib},
however, note also \cite{eqlib2}),
which corresponds to a fast quench from the paramagnetic phase into
the spin glass phase. Then the system is kept at the temperature
$T$ for a first waiting time $t_{w1}$, during which the initial aging takes
place. Note that time is measured in Monte--Carlo seeps through the lattice.
Then the temperature is changed to $T_p$, which is larger
than $T$ in a heat pulse experiment and smaller than $T$ in a
negative temperature cycle experiment. The system is kept at the
temperature $T_p$ for a time $t_p$. Finally the cycle is completed
by changing the temperature back again to the initial temperature $T$,
where the system is aged again for a time $t_{w2}$. After this
time (note that up to this point $\ta=t_{w1}+t_p+t_{w2}$ Monte--Carlo
sweeps have been done).

We performed two different kind of experiments: In the first the whole
simulation is done in zero field, and we stored the
spin configuration at time $\ta$ and measured its overlap with the
spin configurations of the system $t$ Monte--Carlo steps later:
\begin{equation}
C(t,\ta)=\frac1N\sum_i\overline{\av{\sigma_i(t+\ta)\sigma_i(\ta)}}\;.
\label{corr}
\end{equation}
Here $\av{\cdots}$ means a thermal average
(i.e.\ an average over different realizations of the thermal noise,
but the same initial configuration) and the bar means an average over
different realizations of the bond--disorder.

In the second experiment we keep
the system within an external field $h$ during the whole
temperature cycling procedure and switch it off after the the
procedure has been completed (i.e.\ after $\ta$). From that
moment on the thermo-remanent magnetization
\begin{equation}
\mtrm(t,\ta)=\frac1N\sum_i\overline{\av{\sigma_i(t+\ta)}}\;,\label{rem}
\end{equation}
is measured. This field--cooling experiment is exactly what is done
with real spin glasses \cite{Refregier,Lefloch,Granberg,Mattson}.

The linear system size of the samples is L=32 (i.e.\ $\sim3\cdot10^4$
spins), and we averaged over 256 different realizations of the disorder.
There are {\it no} finite size effects observable within
the time scale of 10$^6$ Monte--Carlo steps, which means that the
typical correlation length (or linear domain size within the language
of the droplet picture) is still smaller than half of the linear system
size after $t=10^6$. We believe that our results do not depend significantly
on the choice of the dynamics (\ref{dyn}). Let us adopt the point of
view that spin glasses are critical for all temperatures below the
spin glass transition temperature $T_g$ (note that the correlation
length in the frozen phase is infinite for all temperatures). In this
case we would expect that any microscopic dynamics without order-parameter
conservation (model A in the classification of Hohenberg and Halperin
\cite{HoHa}) will give the same universal results for all temperatures
below $T_g$ as long as the spins are of Ising type and the interactions
are short ranged (so, for instance, also in the case of the short--ranged
Ising spin glass Fe$_{0.5}$Mn$_{0.5}$TiO$_3$ \cite{Gunnarson}.
As soon as one considers e.g.\ Heisenberg spins or
RKKY--interactions the quantitative behavior might change, although
we have nor reason to believe that the qualitative picture of the
results presented here changes significantly.

\section{The correlation function $\bf C(t,\ta)$}

The autocorrelation function $C(t,\ta)$ defined in equation (\ref{corr})
measures the overlap of spin configurations at time $t+\ta$ with that
achieved after aging the system for a time $t_{w1}$ at temperature $T$,
exerting a heat pulse of duration $t_p$ with temperature $T_p$ and
finally aging the system again for a time $t_{w2}$ at temperature $T$.
In figure 1 we choose $T=0.7$, $t_{w1}=10^4$, $t_p=10^2$,
$t_{w2}=10^1(a),10^2(b)$ and various heat pulse temperatures $T_p$.
In figure 1 one observes that the short heat pulse
diminishes the correlations and $C(t,\ta)$ varies smoothly between the two
curves obtained by $T_p=T$ (no heat pulse) and $T_p=\infty$. The latter
curve is identical to that obtained by simple aging with waiting time
$t_{w2}$ since $T_p=\infty$ destroys all correlations grown during the
first waiting time $t_{w1}$. Thus the heat pulse tends to reinitialize
aging, however, not completely as long as $T_p$ is not high enough.
Partial re-initialization has been observed in some experiments
\cite{Granberg,Mattson}. On the other hand, figure 1 is at variance
with other experiments \cite{Refregier,Lefloch}, where aging is
fully reinitialized by applying a heat pulse of temperature only
slightly above $T$ and still significantly below the freezing
temperature $T_g$.

In figure 2 the same parameters as above are used up to the duration of
the pulse, which is now $t_p=10^3$. Note that now the heat pulse is only
one decade shorter than the first waiting time $t_{w1}$ and one observes
differences to figure 1: For $T_p=1.0$ and $1.3$ the correlations are
larger at long times $t$ than those without heat pulse. One possible
interpretation is that on one side the longer heat pulse destroys some
of the correlations originating from the first aging (note that for small
$t$ all curves with $T_p>T$ lie below $T_p=T$), but drives the system
into energetically more favorable states (deeper valleys) like in
simulated annealing \cite{kirk}. Thus it is harder for the system
to escape from the vicinity of the state reached after $\ta$, which
enhances the correlations at long times.

This picture is supported by figure 3, where $T=0.7$, $T_p=1.0$ and
the sum of first waiting time and duration of the heat pulse is kept
constant: $t_{w1}+t_p=1000$. The longer the heat pulse the larger the
correlations $C(t,\ta)$ at large times $t$. For comparison we have
inserted a plot of the function $C(t,t_w=10^5)$ obtained by simple aging
with a much longer waiting time $t_w=10^5\gg\ta$.

This effect is completely absent if one performs a negative temperature
cycling experiment with the same data for $t_{w1}$, $t_p$ and $t_{w2}$.
The result for $T=0.9$ and $T_p=0.7$ (note that now $T_p<T$ is depicted in
figure 4: For increasing duration of the "cold" pulse the correlation
function $C(t,\ta)$ is clearly diminished. It seems that aging is
partially frozen during the negative temperature cycle, the system
cannot reach valleys as deep as those it would explore during simple
aging at temperature $T$ (corresponding to the $t_p=0$ curve). However,
freezing is only partial, since compared with the simple aging curve
$C(t,t_w=t_{w2})$ the correlations are still higher. Let us conclude this
section with this observation of a clear asymmetry
between heat pulse experiments shown in figures 1--3 and negative
temperature cycling experiments shown in figure 4.

\section{The magnetization $\bf M(t,\ta)$}

The correlation function that has been investigated in the last section
is hard to measure in experiments with real spin glasses. Nevertheless
it has a physical meaning and it yields the same information as
the usual susceptibility--measurements in equilibrium (because of the
fluctuation--dissipation theorem, see \cite{JOA,Rieger}) and additional
information in non-equilibrium. In this section we perform the following
procedure already mentioned in section 2, which mimics exactly the
experimental situation described in \cite{Refregier,Granberg,Mattson}:
the temperature cycle is done within a weak external field, by which a
magnetization is induced. After the completion of the cycle the field
is switched off and the decay of the (thermo)-remanent magnetization
(\ref{rem}) is measured.

We show in the following results for rather strong magnetic fields
($h=0.5$), for the simple reason that the data are less scattered,
since the signal (magnetization) is stronger.  We performed also
simulations for $h=0.2$ and $h=0.1$, which give qualitatively the same
results. The differences originating in the fact that $h\sim0.5$
is certainly outside the linear response regime are not observable
on these time scales.

In figure 5 we depicted the results of heat pulses
with various temperatures. The initial aging temperature was $T=0.7$
and $t_{w1}=10^4$, the duration of the pulse was $t_p=10^2$ and the
final waiting time is $t_{w1}=10^1$. For comparison the remanent
magnetization obtained from simple aging at temperature $T=0.7$ and
waiting time $t_w=10$ is shown. One observes that
the heat pulse diminishes the magnetization for times smaller than
$0.1\cdot t_{w1}$, like it does with the correlations in figure 1. The
aging process is again only partially reinitialized, the temperature
of the heat pulse has to be very high nullify the magnetization
obtained during the first waiting time $t_w$. This is of course a
consequence of the high magnetic field, for smaller magnetic fields
the temperatures needed to completely re-initialize the aging process
are significantly smaller.

Furthermore one observes that for $t>0.2\cdot t_{w1}$ the
heat-pulse is able to enhance the magnetization, an effect that becomes
more pronounced the longer the heat pulse is. This effect can be
interpreted in the same way as in the preceding section about the
correlation function $C(t,\ta)$. Here the heat pulse destroys some
of the magnetized domains of the system, however not completely.
Simultaneously it drives the system into energetically more favorable
states, which have a non-vanishing magnetization. After the completion
of the temperature cycle the initial magnetization is smaller, but
it takes longer to escape from this magnetized state and to approach
zero magnetization.

Again, if this picture is correct in essence, one might expect
a different outcome in a negative temperature cycle experiment.
In figure 6 we show such an experiment with $T=0.9$ and cycle temperature
$T_p=0.6$ (note that $T_p<T$ now). The initial waiting time is $t_{w1}=10^3$,
the final waiting time is $t_{w2}=10^2$. For increasing heat pulse length
$t_p$ the remanent magnetization is either unchanged or slightly increased
for all times $t$. Thus the negative temperature cycle destroys nothing
of the magnetization obtained during the initial aging process. The
magnetized domains continue to grow during the cycle (however, at a much
smaller rate), for comparison the magnetization curve obtained for
simple aging at $T=0.9$ with a waiting time $t_w=10^5$ is shown, which
shows a still larger magnetization than that of negative temperature cycling
with $t_p=10^5$.

We conclude that TRM--measurements in temperature cycling experiments
manifests again an asymmetry between heat pulse and negative temperature
cycle experiments and therefore yield the same picture as that obtained
from the calculation of the autocorrelation function $C(t,\ta)$ described
in the last section.

\section{Discussion}

In this section we will summarize and discuss the results
of the numerical experiments presented in this paper. By calculating
the autocorrelation function $C(t,\ta)$ and the thermo-remanent magnetization
$\mtrm(t,\ta)$ we tried to explore the effect of temperature
cycling on the aging process within the three--dimensional Ising spin
glass model. We demonstrated that by a heat pulse, which is short compared to
the initial waiting time, the aging process is partially re-initialized.
On the other side, a negative temperature cycle experiment partially
freezes the system into the (domain)--state reached during the initial
aging process. Some experiments on real spin glasses show a much clearer
outcome \cite{Refregier,Lefloch}, which, nevertheless, might be interpreted
to concur with our observation. And finally --- as pointed out in
\cite{Lefloch} --- this asymmetry would pledge in favor of the hierarchical
picture mentioned in the introduction and against the droplet picture.

However, our results are on a very qualitative level and the dynamical
processes involved are still microscopic on a logarithmic time scale.
Thus it might be hard to verify one phenomenological, macroscopic theory
and falsify another on the ground of our numerical data, although they
have been obtained by state--of--the--art simulations with the most
efficient existing algorithm implemented on one of the fastest computers
of today. The parameter space for this kind of experiments is
essentially six--dimensional ($T$, $T_p$, $t_{w1}$, $t_p$, $t_{w2}$ and $h$),
therefore a systematic investigation, as was done for simple aging experiments
in the same model \cite{Rieger}, seems to be forbidden. Therefore we had to
confine ourselves to demonstrate what kind of scenario for temperature cycling
experiments is obtained for the model, time scales and quantities under
considerations and can only offer a possibly speculative interpretation.

Furthermore we would like to point out that very strong crossover phenomena
are observable within our results as soon as the duration of the heat pulse
of the final waiting time become comparable to the initial aging time.
We interpreted them within a picture of a relaxation in a rough free energy
landscape, which again seems to be most appropriate for the results we
obtained. This picture is rather flexible and is able to explain a lot of
features in a frustrated system --- real or theoretical, and regardless of the
existence of a phase transition. One of the phenomena that are fully
neglected in a theory that is based on the assumption of a relaxation
within a complicated free energy landscape is that of the growth of
spatial correlations during the aging process. They are on the other side
the basic ingredience of the droplet model \cite{FiHu}. Although both
theories seem to make contradicting predictions (e.g.\ the symmetry
or asymmetry of heat pulse and negative temperature-cycling experiments)
our impression is that they have more in common than usually admitted.

We think that it might be very useful to try to find a synthesis of
both models, not only in order to be able to describe the growth of spatial
correlations and their destruction by a heat pulse and their freezing
during a negative temperature cycle. Domain growth has not been investigated
by direct measurements up to now (for experiments that investigate this
matter indirectly see \cite{Mattson,Granberg2,Schins}). However, in numerical
simulations one
has an immediate access to the quantities of interest and work on this
subject is in progress\cite{Rieger2}. It is our impression, obtained from
the results presented in this paper and in other publications
\cite{JOA,Rieger}, that the simulation of finite dimensional Ising spin
glass models can make relevant predictions for real spin glasses, too,
and will prove to be a very useful tool in testing and improving
phenomenological theories for them.

\section[*]{Acknowledgement}

We would like to thank the HLRZ at the research center in J\"ulich for
the generous allocation of computing time (approximately 250 CPU hours)
on the Cray YMP. This work was performed within the SFB 341
K\"oln--Aachen--J\"ulich.

\begin{figcap}
\item \label{fig1}
The spin autocorrelation function $C(t,\ta)$ defined in (\ref{corr})
in dependence of $t$ (number of MC--steps)
for various heat pulse temperatures $T_p$. The
other temperature cycling parameters are $T=0.7$, $t_{w1}=10^4$, $t_p=10^2$
and $t_{w2}=10^1$ in the upper figure and $t_{w2}=10^2$ in the lower figure.
 From top to bottom it is  $T_p=T$ ($\circ$), $T_p=1.0$, $1.3$, $1.6$, $1.9$,
$2.2$, $2.5$ (full lines) and $T_p=\infty$ ($\Box$).
The error bars are significantly smaller than the symbols for the curves
plotted with points.

\item \label{fig2}
Same as in figure 1, but with a longer duration of the heat pulse
$t_p=10^3$ and pulse temperatures (from top to bottom at $t\sim10^3$)
$T_p=T$ ($\circ$), $T_p=1.0$, $1.3$, $1.6$, $2.0$ (full lines)
and $T_p=\infty$ ($\Box$).

\item \label{fig3}
$C(t,\ta)$ in dependence of $t$ with $T=0.7$, $T_p=1.0$ and $t_{w2}=100$.
The sum $t_{w1}+t_p=1000$ is constant, from bottom to top it is
$t_p=0$ ($\diamond$), $t_p=10$, $100$, $300$, $500$ and $900$ (full lines).
The top curve ($\Box$) is just for comparison --- it is the function
$C(t,t_w=10^5)$ obtained from simple aging with waiting time $t_w=10^5\gg\ta$.

\item \label{fig4}
$C(t,\ta)$ from a negative temperature cycling experiment with
$T=0.9$, $T_p=0.7$ ($<T$!) and $t_{w2}=100$.
As in figure 3 the sum $t_{w1}+t_p=1000$ is constant, from top to bottom
it is $t_p=0$ ($\diamond$), $t_p=10$, $100$, $500$ and $900$ (full lines).
The bottom curve ($\Box$) is just for comparison --- it is the function
$C(t,t_w=10^2)$ obtained from simple aging with waiting time
$t_w=10^2=t_{w2}$.

\item \label{fig5}
The thermo-remanent magnetization $\mtrm(t,\ta)$ defined in (\ref{rem})
in dependence of $t$ (number of MC--steps) for various heat pulse
temperatures $T_p$. The other temperature cycling parameters are $T=0.7$,
$t_{w1}=10^4$, $t_p=10^2$ and $t_{w2}=10^1$.
 From top to bottom (at $t=100$) it is  $T_p=T$ ($\diamond$),
$T_p=1.0$, $1.3$, $1.6$, $1.9$, $2.2$, $2.5$ (full lines) and
$T_p=\infty$ ($\Box$). The error bars are significantly smaller than
the symbols for the curves plotted with points.

\item \label{fig6}
$\mtrm(t,\ta)$ from a negative temperature cycling experiment with
$T=0.9$, $T_p=0.6$ ($<T$!), $t_{w1}=10^3$ and $t_{w2}=10^2$.
 From bottom to top (at $t=1000$) it is $t_p=0$ ($\Box$),
$t_p=10^2$, $10^3$, $10^4$ and $10^5$. The top curve ($\diamond$)
is just for comparison --- it is the function $\mtrm(t,t_w=10^5)$
obtained from simple aging with waiting time  $t_w=10^5$.

\end{figcap}

\end{document}